\documentclass[final,5p,times,twocolumn]{elsarticle}

\usepackage[T1]{fontenc}
\usepackage[utf8]{inputenc}
\usepackage{amsmath,amssymb,amsfonts,mathtools,bm}
\usepackage{graphicx}
\usepackage{microtype}
\usepackage{xcolor}
\usepackage{hyperref}
\hypersetup{colorlinks=true,linkcolor=blue,citecolor=blue,urlcolor=blue}
\allowdisplaybreaks
\biboptions{sort&compress}

\newcommand{\dd}{\mathrm{d}}
\newcommand{\cT}{\mathcal{T}}
\newcommand{\cA}{\mathcal{A}}
\newcommand{\Order}{\mathcal{O}}

\journal{Physics Letters B}

\begin{document}
\begin{frontmatter}

\title{Regular black holes from dynamical-tension string hedgehogs}
\author[ibs]{Sebastian Bahamonde}
\ead{sbahamondebeltran@gmail.com}
\address[ibs]{Cosmology, Gravity, and Astroparticle Physics Group, Center for Theoretical Physics of the Universe, Institute for Basic Science (IBS), Daejeon 34126, Korea}

\author[bgu,fias,basic]{Eduardo I. Guendelman}
\ead{eduardoleonguendelman@gmail.com}
\address[bgu]{Department of Physics, Ben-Gurion University of the Negev, Beer-Sheva, Israel}
\address[fias]{Frankfurt Institute for Advanced Studies (FIAS), Ruth-Moufang-Strasse 1, 60438 Frankfurt am Main, Germany}
\address[basic]{Bahamas Advanced Study Institute and Conferences, 4A Ocean Heights, Hill View Circle, Stella Maris, Long Island, The Bahamas}

\begin{abstract}
A spherical cloud of radial strings with non-zero constant tension has an energy density proportional to $r^{-2}$ and therefore produces a singular conical geometry rather than an asymptotically flat regular black hole. We show that this obstruction can be removed within the modified-measure formulation of strings. A continuum of radial worldsheets whose tensions are generated by a composite bulk tension scalar reduces, on a constrained $\mathrm{SO}(3)$ hedgehog branch, to Einstein gravity coupled to a nonlinear function $K(Y)$ and an auxiliary three-form. This construction gives model-independent conditions directly in terms of the radial tension $\mathcal{T}(r)$. Finite curvature at the centre requires $\mathcal{T}=\mathcal{O}(r^2)$, while the null energy condition forbids a faster leading power. Thus a positive regular centre satisfying the null energy condition is necessarily de Sitter. Finite ADM mass requires the tension to be screened at infinity, whereas the dominant energy condition cannot hold globally for a non-trivial positive profile screened faster than $r^{-2}$. An exact cubic profile realises these properties and gives a continuous family of asymptotically flat geometrically regular black holes whose first correction to Schwarzschild appears at order $r^{-4}$.
\end{abstract}

\begin{keyword}
regular black holes \sep dynamical string tension \sep string clouds \sep hedgehog scalar fields
\end{keyword}

\end{frontmatter}

\section{Introduction}

Regular black holes replace the central curvature singularity by a finite-curvature region while retaining an event horizon and the standard far field. Many such geometries are known, either as effective metrics or as exact solutions supported by nonlinear matter sectors, but identifying a simple source with a clear physical interpretation remains a non-trivial problem; see Refs.~\cite{Lan:2023cvz,CarballoRubio:2025rnm} for recent reviews. In the static spherically symmetric case, a de Sitter core is a particularly common mechanism, as in the Hayward geometry~\cite{Hayward:2005gi}. The price is an anisotropic source whose microscopic meaning is often less transparent than the resulting metric.

A natural anisotropic system is a spherical distribution of radially oriented strings. The standard cloud of strings introduced by Letelier~\cite{Letelier:1979ej}, and its fluid generalisations~\cite{Stachel:1980fg}, have a distinguished longitudinal equation of state. However, a constant-tension radial string cloud behaves as $\rho\propto r^{-2}$. It therefore produces the same conical large-distance structure encountered for global monopoles~\cite{Barriola:1989hx}, and is singular at the centre. Hedgehog sources in gravitational settings were also considered in connection with vacuum-bubble evolution~\cite{Guendelman:1991qb}, and gravitating string hedgehogs with the same constant-tension behaviour have been studied explicitly~\cite{Kawai:2010zh}. More recent constructions have regularised string fluids by prescribing a radial equation of state, a new effective string source, or a screening deformation of the geometry~\cite{Santos:2025yjr,Muniz:2025lsc,Estrada:2026rfs}. String-fluid configurations have also appeared recently in Born--Infeld models motivated by open-string tachyon condensation, where radial electric-flux tubes, equivalently open strings, provide the source for rotating Kerr black-hole mimickers~\cite{Brustein:2024yxu}. These results show that regular or ultracompact string-supported geometries can be constructed, but they do not derive the screened radial tension required here from a modified-measure world-sheet mechanism.

In modified-measure string theory the tension is not inserted as a fixed parameter. It appears as an integration constant of an auxiliary world-sheet gauge equation~\cite{Guendelman:2000ij,Guendelman:2002qs}; related dynamical-tension mechanisms were discussed in Ref.~\cite{Townsend:1991xk}. A conserved current induced by a bulk scalar can further turn the tension into a position-dependent quantity~\cite{Ansoldi:2005fs}. This provides a natural setting in which the failure of the constant-tension hedgehog may be addressed without postulating its radial profile directly at the level of the stress tensor.

In this Letter, we establish a reduced-action correspondence between an isotropic continuum of modified-measure radial strings and a constrained scalar hedgehog coupled to an auxiliary three-form. The world-sheet integration constant gives precisely the singular Letelier term, so regularity and asymptotic flatness select the branch on which it vanishes. The remaining tension is generated by a composite bulk scalar and is controlled in amplitude by the three-form integration constant. We then derive universal relations between the logarithmic slope of the tension, regularity and the energy conditions. In particular, regularity together with the null energy condition fixes the leading central behaviour to $\cT\propto r^2$, while a sufficiently fast asymptotic screening necessarily violates the dominant energy condition somewhere. Finally, we apply the construction to the exact hedgehog family found in Ref.~\cite{Bahamonde:2026ghh}. The geometry itself was obtained there; its world-sheet origin and the general tension constraints derived below are new.

\section{Dynamical-tension string hedgehog}

We first recall the minimal world-sheet mechanism needed in the following. For each string, let
\begin{align}
 S_i={}&-\int \dd^2\sigma\,\Omega_i(\varphi)
 \left[\frac{1}{2}\gamma_i^{ab}h^{(i)}_{ab}
 -\frac{\epsilon^{ab}}{2\sqrt{-\gamma_i}}F^{(i)}_{ab}\right]
 -\int \dd^2\sigma\,a^{(i)}_a j_i^a ,
 \label{eq:worldsheet-action}
\end{align}
where $h^{(i)}_{ab}=g_{\mu\nu}\partial_aX_i^\mu\partial_bX_i^\nu$ is the induced metric,
$F^{(i)}_{ab}=2\partial_{[a}a^{(i)}_{b]}$, and
\begin{equation}
 \Omega_i(\varphi)=\frac{1}{2}\epsilon_{mn}\epsilon^{ab}
 \partial_a\varphi_i^m\partial_b\varphi_i^n
 \label{eq:modified-measure}
\end{equation}
is a metric-independent measure density. We shall take the tension scalar to be the composite bulk scalar $\phi(C,Y)$. The corresponding identically conserved world-sheet current is
\begin{equation}
 j_i^a=e\epsilon^{ab}\partial_b\phi(C,Y)\big|_{X_i} .
 \label{eq:worldsheet-current}
\end{equation}
With the sign convention in Eq.~\eqref{eq:worldsheet-action}, variation with respect to $a^{(i)}_a$ gives
\begin{equation}
 \epsilon^{ab}\partial_b\left(\frac{\Omega_i}{\sqrt{-\gamma_i}}\right)=j_i^a,
 \qquad
 \cT_i\equiv\frac{\Omega_i}{\sqrt{-\gamma_i}}
 =T_i+e\phi(C,Y) \,,
 \label{eq:dynamical-tension}
\end{equation}
where $e$ is the coupling of the world-sheet gauge field to the bulk scalar.
Thus $T_i$ is a world-sheet integration constant, while $\phi$ generates a variable contribution. Up to a boundary term, the gauge-field sector is $\frac12\int(\Omega_i/\sqrt{-\gamma_i}-e\phi)\epsilon^{ab}F^{(i)}_{ab}$. Using Eq.~\eqref{eq:dynamical-tension}, the embedding action becomes
\begin{equation}
 \begin{aligned}
 S_i^{\rm eff}={}&-\frac12\int\dd^2\sigma\sqrt{-\gamma_i}\,
 \cT_i(X_i)\gamma_i^{ab}h^{(i)}_{ab}\\
 &\longrightarrow-\int\dd^2\sigma\,\cT_i(X_i)\sqrt{-h_i}.
 \end{aligned}
 \label{eq:effective-NG}
\end{equation}
where the last step eliminates the intrinsic metric. The radial profile is therefore fixed by a world-sheet gauge equation rather than assigned directly to an effective stress tensor~\cite{Guendelman:2000ij,Guendelman:2002qs,Ansoldi:2005fs}.

The required bulk fields can be introduced through the parent action
\begin{equation}
 \begin{aligned}
 S_{\rm p}={}&\int \dd^4x\sqrt{-g}\left[
 \frac{R}{16\pi G}
 +\lambda\left(\delta_{IJ}\Phi^I\Phi^J-\eta^2\right)\right]\\
 &-\frac{1}{3!}\int \dd^4x\,
 \varepsilon^{\mu\nu\rho\sigma}\cA_{\mu\nu\rho}\partial_\sigma C
 +\sum_i S_i .
 \end{aligned}
 \label{eq:parent-action}
\end{equation}
The three-form $\cA_{\mu\nu\rho}$ carries no local propagating degree of freedom in four dimensions and generates a constant on shell~\cite{Aurilia:1980xj,Henneaux:1989zc}. We have also defined
\begin{equation}
 Y=\frac{1}{2}\delta_{IJ}\nabla_\mu\Phi^I\nabla^\mu\Phi^J .
 \label{eq:Y-def}
\end{equation}
The fixed-norm triplet admits the hedgehog branch
\begin{equation}
 \Phi^I=\eta n^I(\theta,\varphi),
 \qquad
 n^I=(\sin\theta\cos\varphi,\sin\theta\sin\varphi,\cos\theta),
 \label{eq:hedgehog}
\end{equation}
which is spherically symmetric up to a compensating internal rotation, as in sigma-model and Skyrme hedgehogs~\cite{Canfora:2013osa}.

To keep the reduction independent of the areal gauge, consider
\begin{equation}
 \dd s^2=-A(r)\dd t^2+\frac{\dd r^2}{B(r)}+\mathcal R(r)^2\dd\Omega^2,
 \qquad
 Y=\frac{\eta^2}{\mathcal R(r)^2}.
 \label{eq:general-metric}
\end{equation}
The embedding equation following from Eq.~\eqref{eq:effective-NG} can be written as $\cT K^\mu+\perp^{\mu\nu}\nabla_\nu\cT=0$, where $K^\mu$ is the mean-curvature vector and $\perp^{\mu\nu}$ projects orthogonally to the world sheet. A surface at fixed angular position is totally geodesic in Eq.~\eqref{eq:general-metric}, while $\nabla_\mu\cT$ is radial and therefore tangent to it. The radial worldsheets are consequently exact embeddings also when their tension varies. Using $(t,r)$ as world-sheet coordinates, their induced volume element is $\sqrt{A/B}\,\dd t\dd r$. Smearing the strings uniformly over the sphere with angular normalisation $N_{\rm s}$, namely $\dd N=N_{\rm s}\dd\Omega/(4\pi)$, gives
\begin{equation}
 S_{\rm hed}=-\frac{N_{\rm s}}{4\pi}\int \dd t\dd r\dd\Omega
 \sqrt{\frac{A}{B}}\left[T_0+e\phi(C,Y)\right],
 \label{eq:string-continuum}
\end{equation}
where spherical symmetry requires a common value, or at least an angle-independent average, of the constants $T_i$.

We now choose the tension field according to
\begin{equation}
 e\phi(C,Y)=\frac{4\pi\eta^2}{N_{\rm s}}\frac{C\widehat K(Y)}{Y}.
 \label{eq:tension-field-choice}
\end{equation}
Using $\mathcal R^2=\eta^2/Y$, Eq.~\eqref{eq:string-continuum} becomes
\begin{equation}
 S_{\rm hed}=-\int \dd^4x\sqrt{-g}\left[
 C\widehat K(Y)+\frac{N_{\rm s}T_0}{4\pi\eta^2}Y\right].
 \label{eq:reduced-string-action}
\end{equation}
Consequently, on the constrained spherical hedgehog branch, the parent string system reduces to the scalar action used in Ref.~\cite{Bahamonde:2026ghh}, supplemented by a term linear in $Y$. This equality holds at the level of the reduced action and its spherically symmetric variations. It is not an off-shell identification between a microscopic string ensemble and the complete triplet theory away from this sector.

Variation of the three-form gives
\begin{equation}
 \partial_\mu C=0,
 \qquad C=\rho_0,
 \label{eq:C-constant}
\end{equation}
so the amplitude is an integration constant rather than a coupling. The effective kinetic function is
\begin{equation}
 K(Y)=\rho_0\widehat K(Y)+\frac{N_{\rm s}T_0}{4\pi\eta^2}Y .
 \label{eq:effective-K}
\end{equation}
In the areal gauge $\mathcal R=r$, its stress tensor is
\begin{equation}
 \begin{gathered}
 T^\mu{}_{\nu}=\mathrm{diag}(-\rho,p_r,p_t,p_t),\\
 \rho=K,\qquad p_r=-\rho,\qquad p_t=-K+YK_Y .
 \end{gathered}
 \label{eq:stress-K}
\end{equation}
Equivalently, defining the common radial tension
\begin{equation}
 \cT(r)=T_0+\frac{4\pi r^2\rho_0}{N_{\rm s}}\widehat K\left(\frac{\eta^2}{r^2}\right),
 \label{eq:tension-reconstruction}
\end{equation}
we obtain the exact relations
\begin{equation}
 \rho=\frac{N_{\rm s}}{4\pi r^2}\cT,
 \qquad
 p_r=-\rho,
 \qquad
 p_t=-\rho-\frac{r}{2}\rho'
 =-\frac{N_{\rm s}}{8\pi r}\cT'.
 \label{eq:string-stress}
\end{equation}
The last equality is the transverse stress required by conservation. A prescribed variable tension carried only by otherwise free radial Nambu--Goto strings would not be separately conserved. Here it is a composite bulk quantity, and its $Y$ dependence generates the required tangential pressure through the reduced action.

Since $T^t{}_t=T^r{}_r$, the metric can be written, after a constant rescaling of time, as
\begin{equation}
 \dd s^2=-f(r)\dd t^2+\frac{\dd r^2}{f(r)}+r^2\dd\Omega^2,
 \qquad
 f(r)=1-\frac{2Gm(r)}{r}.
 \label{eq:metric-mass}
\end{equation}
Einstein's equations reduce to
\begin{equation}
 m'(r)=4\pi r^2\rho(r)=N_{\rm s}\cT(r),
 \qquad
 M=N_{\rm s}\int_0^\infty \cT(r)\dd r,
 \label{eq:mass-tension}
\end{equation}
where $m(0)=0$ has been imposed and the second relation applies when the integral converges. The ADM mass is therefore the integrated radial tension of the hedgehog.

\section{Regularity and energy conditions}

The previous relations allow the relevant restrictions to be stated without choosing a particular model. For positive tension, define its logarithmic slope
\begin{equation}
 \alpha(r)=\frac{\dd\ln\cT}{\dd\ln r}.
 \label{eq:alpha-def}
\end{equation}
Equation~\eqref{eq:string-stress} gives
\begin{equation}
 w_t\equiv\frac{p_t}{\rho}=-\frac{\alpha}{2}.
 \label{eq:wt-alpha}
\end{equation}
The radial null energy condition is saturated. The tangential null energy condition, the dominant energy condition and, once the null energy condition is imposed, the strong energy condition become, respectively,
\begin{equation}
 \mathrm{NEC}:\ \alpha\leq2,
 \qquad
 \mathrm{DEC}:\ -2\leq\alpha\leq2,
 \qquad
 \mathrm{SEC}:\ \alpha\leq0.
 \label{eq:energy-conditions}
\end{equation}
Thus the energy conditions are controlled only by the local rate at which the string tension changes.

Suppose that near the centre
\begin{equation}
 \cT(r)=\tau_s r^s+\Order(r^s),
 \qquad \tau_s>0.
 \label{eq:central-power}
\end{equation}
Finite density and finite curvature require $s\geq2$. On the other hand, the null energy condition requires $s\leq2$. Hence
\begin{equation}
 \boxed{\quad \text{regularity}+\mathrm{NEC}\quad\Longrightarrow\quad
 \cT(r)=\tau_2r^2+\Order(r^2)\quad}
 \label{eq:central-selection}
\end{equation}
for a non-vanishing positive central source. In this case
\begin{equation}
 \begin{aligned}
 \rho(0)&=\frac{N_{\rm s}\tau_2}{4\pi},
&p_r(0)&=p_t(0)=-\rho(0),\\
 f(r)&=1-\frac{2GN_{\rm s}\tau_2}{3}r^2+\Order(r^2).
 \end{aligned}
 \label{eq:desitter-core}
\end{equation}
Therefore, the de Sitter core is not an independent assumption: it follows from regularity and the null energy condition for the radial-string equation of state. A faster vanishing tension, $s>2$, gives a Minkowski-like centre but violates the tangential null energy condition arbitrarily close to it. The same central behaviour supplies a natural world-sheet boundary condition. For a semi-infinite radial string, the endpoint momentum obtained from Eq.~\eqref{eq:effective-NG} is
\begin{equation}
 \Pi^r{}_{\mu}\equiv
 -\cT\sqrt{-h}\,h^{rb}g_{\mu\nu}\partial_bX^\nu .
 \label{eq:endpoint-momentum}
\end{equation}
For the radial embedding in Eq.~\eqref{eq:metric-mass},
$|\Pi^r{}_{r}|=\cT$. Hence Eq.~\eqref{eq:central-selection} removes the endpoint force at the regular centre, while asymptotic screening removes it at infinity. No supporting central membrane is required at the level of the background, in contrast with constant-tension string--membrane hedgehogs~\cite{Kawai:2010zh}.

The constant part of Eq.~\eqref{eq:dynamical-tension} is now seen to be obstructed. For $\cT=T_0\neq0$, and in the absence of a point-mass contribution,
\begin{equation}
 \begin{aligned}
 \rho&=\frac{N_{\rm s}T_0}{4\pi r^2},
 &m&=N_{\rm s}T_0r,\\
 f&=1-2GN_{\rm s}T_0,
 &R&=\frac{4GN_{\rm s}T_0}{r^2}.
 \end{aligned}
 \label{eq:constant-tension-obstruction}
\end{equation}
The centre is singular and the asymptotic geometry is conical rather than flat. Adding a Schwarzschild integration constant cannot regularise it. Therefore a regular asymptotically flat positive-tension ensemble requires $T_0=0$; if every individual $T_i$ is non-negative, each world-sheet integration constant must vanish. The non-zero mass is nevertheless retained because the bulk tension field in Eq.~\eqref{eq:tension-field-choice} is multiplied by the independent three-form constant $\rho_0$.

Finite ADM mass further requires $\cT$ to be integrable and hence screened at large radius. A continuous positive profile satisfying Eq.~\eqref{eq:central-selection} must then possess at least one maximum. At such a point $\alpha=0$, so $p_t=0$ and the strong energy condition changes character. There is also a useful asymptotic restriction. If the dominant energy condition held everywhere, its lower bound in Eq.~\eqref{eq:energy-conditions} would imply
\begin{equation}
 \frac{\dd}{\dd r}\left(r^2\cT\right)
 =r\cT(2+\alpha)\geq0.
 \label{eq:DEC-monotonicity}
\end{equation}
A non-trivial positive profile satisfying the dominant energy condition therefore cannot be screened faster than $r^{-2}$. For a power-law tail,
\begin{equation}
 \cT(r)=\tau_\infty r^{-q}+\Order(r^{-q}),
 \qquad q>1,
 \label{eq:tail-power}
\end{equation}
finite mass requires $q>1$, while
\begin{equation}
 f(r)=1-\frac{2GM}{r}
 +\frac{2GN_{\rm s}\tau_\infty}{q-1}\,r^{-q}+\Order(r^{-q}),
 \qquad
 w_t\longrightarrow\frac{q}{2}.
 \label{eq:tail-metric}
\end{equation}
Hence the dominant energy condition allows $q\leq2$. A matter correction smaller than the Reissner--Nordstr\"om order $r^{-2}$ requires $q>2$ and necessarily violates the dominant energy condition in the asymptotic region. This is a structural trade-off of the radial-string description, independent of the particular profile chosen below.

\section{Exact screened branch}

A simple family satisfying the central condition is generated by
\begin{equation}
 \widehat K_n(Y)=\left(\frac{Y}{Y+\mu_\star^2}\right)^n,
 \qquad
 L=\frac{\eta}{\mu_\star}.
 \label{eq:Kn}
\end{equation}
The corresponding bulk tension field in Eq.~\eqref{eq:tension-field-choice} is
\begin{equation}
 e\phi_n(C,Y)=\frac{4\pi\eta^2 C}{N_{\rm s}}
 \frac{Y^{n-1}}{(Y+\mu_\star^2)^n}.
 \label{eq:phi-n}
\end{equation}
For $n>1$, it vanishes both for $Y\rightarrow\infty$ and $Y\rightarrow0$, namely at the centre and at infinity on the hedgehog branch. Setting $T_0=0$ and using $C=\rho_0$, one obtains
\begin{equation}
 \rho_n(r)=\frac{\rho_0}{(1+x^2)^n},
 \qquad
 \cT_n(r)=\frac{4\pi\rho_0L^2}{N_{\rm s}}
 \frac{x^2}{(1+x^2)^n},
 \qquad x=\frac{r}{L}.
 \label{eq:rho-T-n}
\end{equation}
The tangential equation of state and the mass are
\begin{align}
 w_t(x)&=-1+\frac{nx^2}{1+x^2},
 \label{eq:wt-n}\\
 m_n(r)&=4\pi\rho_0L^3\int_0^x\frac{u^2\dd u}{(1+u^2)^n},
 \qquad
 M_n=\pi^{3/2}\rho_0L^3\frac{\Gamma(n-3/2)}{\Gamma(n)},
 \label{eq:mass-n}
\end{align}
where $n>3/2$ is required for finite ADM mass. The weak energy condition holds for every $n>0$. Within the finite-mass family, the dominant energy condition holds globally only for $n\leq2$. Thus $n=2$ is the unique positive integer member with finite mass and global dominant energy condition, but its metric contains an $r^{-2}$ correction. The cubic case $n=3$ is the first integer member whose matter correction decays faster; its dominant energy condition is violated only for $x>\sqrt{2}$.

\begin{figure}[t]
 \centering
 \includegraphics[width=1\linewidth]{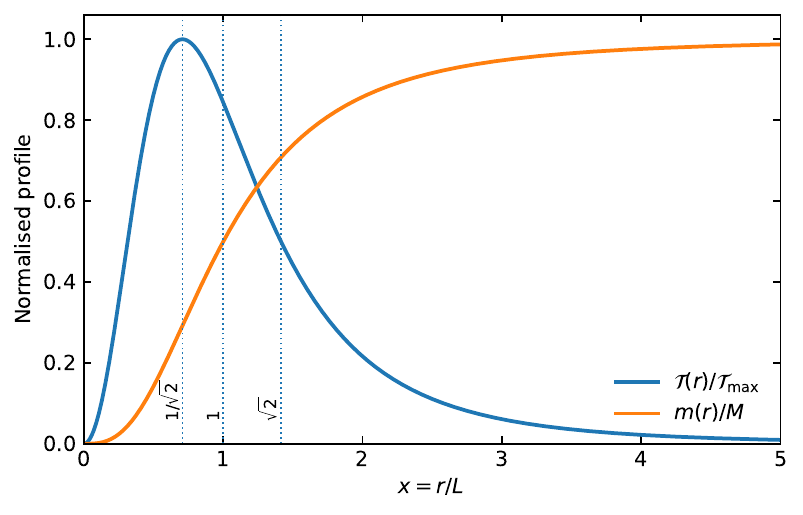}
 \caption{Generated tension and enclosed mass for the cubic branch. The vertical lines at $x=1/\sqrt{2}$, $x=1$ and $x=\sqrt{2}$ mark, respectively, the tension maximum and $p_t=0$, the half-mass radius, and the boundary of the dominant energy condition.}
 \label{fig:tension-mass}
\end{figure}

For $n=3$, the exact mass function and ADM mass are~\cite{Bahamonde:2026ghh}
\begin{equation}
 m(r)=\frac{\pi\rho_0L^3}{2}
 \left[\arctan x+\frac{x(x^2-1)}{(1+x^2)^2}\right],
 \qquad
 M=\frac{\pi^2}{4}\rho_0L^3.
 \label{eq:m3}
\end{equation}
The metric function is therefore
\begin{equation}
 f(r)=1-\frac{4GM}{\pi r}
 \left[\arctan x+\frac{x(x^2-1)}{(1+x^2)^2}\right].
 \label{eq:f3}
\end{equation}
Its two relevant limits are
\begin{align}
 f(r)&=1-\frac{8\pi G\rho_0}{3}r^2+\Order(r^4),
 &&r\rightarrow0,
 \label{eq:f3-small}\\
 f(r)&=1-\frac{2GM}{r}+\frac{32GML^3}{3\pi r^4}
 +\Order(r^{-6}),
 &&r\rightarrow\infty.
 \label{eq:f3-large}
\end{align}
The generated tension reaches its maximum at $x=1/\sqrt{2}$,
\begin{equation}
 \cT_{\rm max}=\frac{16\pi\rho_0L^2}{27N_{\rm s}}
 =\frac{64M}{27\pi N_{\rm s}L},
 \label{eq:Tmax}
\end{equation}
where $p_t=0$ and the strong energy condition starts to be satisfied. Moreover, $m(L)=M/2$, while the dominant energy condition ceases to hold at $x=\sqrt{2}$. These scales are displayed in Fig.~\ref{fig:tension-mass}.

For completeness, defining $\chi=GM/L$ and
\begin{equation}
 F(x)=\arctan x+\frac{x(x^2-1)}{(1+x^2)^2},
 \label{eq:F-def}
\end{equation}
the extremal configuration satisfies $xF'(x)=F(x)$. The solution is horizonless for $\chi<\chi_{\rm ext}$, extremal at equality, and has two horizons for $\chi>\chi_{\rm ext}$. The exact family is continuous within one fixed theory because $M$ varies with the three-form integration constant $\rho_0$ rather than with a coupling.

\section{Conclusions}

We have shown that the regular hedgehog source of Ref.~\cite{Bahamonde:2026ghh} admits a direct origin from dynamical-tension worldsheets. On the static spherical branch, an isotropic continuum of modified-measure radial strings reduces to a nonlinear constrained-triplet action. The ordinary world-sheet integration constant produces exactly the linear-$Y$ string-cloud sector. Its non-zero value gives a $1/r^2$ density, a conical asymptotic geometry and a curvature singularity. Regularity and asymptotic flatness therefore select the branch with vanishing constant tension. The remaining tension is generated by a bulk field, while the auxiliary three-form fixes its amplitude as the integration constant $\rho_0$. In particular, the three-form does not generate the radial profile by itself; that profile follows from the $Y$ dependence of the tension field.

The tension language also makes several general properties immediate. For a positive radial-string hedgehog, regularity and the null energy condition force $\cT\sim r^2$ and hence a de Sitter centre; the same behaviour makes the central endpoint force vanish. Finite mass requires the tension to rise from zero, attain a maximum and be screened at infinity. The dominant energy condition prevents a non-trivial profile from decaying faster than $r^{-2}$ everywhere, so removing an $r^{-2}$ matter correction entails a violation of this condition in some region. The exact cubic branch illustrates the trade-off cleanly: it has a de Sitter core, finite ADM mass and an $r^{-4}$ leading correction to Schwarzschild, while the dominant energy condition fails only beyond $r=\sqrt{2}L$.

The correspondence established here is deliberately restricted to the constrained spherically symmetric sector. It does not provide a microscopic completion of the continuum string distribution, and it does not address perturbative stability. Moreover, as in Ref.~\cite{Bahamonde:2026ghh}, regularity is geometric: the fixed-modulus hedgehog order parameter is not smooth at the central point even though the stress tensor and curvature invariants are finite. A fully covariant continuum formulation and its perturbations would be required to determine whether the dynamical-tension picture remains valid beyond the exact backgrounds considered here.
\section*{Acknowledgements}
S.B. is supported by the Institute for Basic Science under the project code IBS-R018-D3. 
E.I.G. is grateful to COSMOVERSE, COST Action CA21136, COST Action CA23130 -- Bridging high and low energies in search of quantum gravity (BridgeQG), and Ben-Gurion University of the Negev for generous support.
\bibliographystyle{elsarticle-num}
\bibliography{references}

\end{document}